  \documentclass[prb, twocolumn,final,aps,superscriptaddress,eqsecnum,10pt
  amssymb,amsmath,amsfonts,showpacs,floatfix,citeautoscript]{revtex4-1}
\usepackage{amsmath}
\usepackage{natbib}
\usepackage[pdftex]{graphicx}
\usepackage{epstopdf}
\usepackage{times}
\usepackage{txfonts}
\usepackage{textcomp}
\usepackage{amssymb}
\usepackage{footmisc}
\usepackage[colorlinks, bookmarks=false, citecolor=blue, linkcolor=red,
urlcolor=blue]{hyperref}
\def\be {\begin{equation}}
\def\ee {\end{equation}}

\def\horparallel{ \lower.5ex\hbox{ \includegraphics[width=2ex]{fig-hor.pdf}}\,\,
}
\def\vertparallel{ \lower.5ex\hbox{
\includegraphics[width=2ex]{fig-vert.pdf}}\,\, }
\def\gsim{\mathrel{\rlap{\lower4pt\hbox{\hskip1pt$\sim$}}\raise1pt\hbox{$>$}}}
\def\lsim{\mathrel{\rlap{\lower4pt\hbox{\hskip1pt$\sim$}}\raise1pt\hbox{$<$}}}
\newcommand{\dagga}{{\phantom{\dagger}}}

\begin{document}

\title{Semiclassical theory of the magnetization process of the triangular lattice Heisenberg model}

\author{Tommaso Coletta}
\affiliation{School of Engineering, University of Applied Sciences of Western Switzerland (HES-SO), CH-1950 Sion, Switzerland} 
\author{Tam\'as A. T\'oth}
\affiliation{Haute \'ecole de gestion de Gen\`eve, University of Applied Sciences of Western Switzerland (HES-SO), CH-1227 Carouge , Switzerland}
\author{Karlo Penc}
\affiliation{Institute for Solid State Physics and Optics, Wigner Research
Centre for Physics, Hungarian Academy of Sciences, H-1525 Budapest, P.O.B. 49, Hungary}
\affiliation{MTA-BME Lend\"ulet Magneto-optical Spectroscopy Research Group, 1111
Budapest, Hungary}
\author{Fr\'ed\'eric Mila}
\affiliation{Institute of Physics, \'Ecole Polytechnique F\'ed\'erale de Lausanne
   (EPFL), CH-1015 Lausanne, Switzerland}
\date{\today}

\begin{abstract}
Motivated by the numerous examples of 1/3 magnetization plateaux in the triangular lattice Heisenberg antiferromagnet
with spins ranging from 1/2 to 5/2, we revisit the semiclassical calculation of the magnetization curve of
that model, with the aim of coming up with a simple method that allows one to calculate the full magnetization curve, 
and not just the critical fields of the 1/3 plateau. We show that it is actually possible to calculate the magnetization curve including the first 
quantum corrections and the appearance of the 1/3 plateau entirely within linear spin-wave theory, with
predictions for the critical fields that agree to order 1/S with those derived a long-time ago on the basis
of arguments that required to go beyond linear spin-wave theory. This calculation relies on the 
central observation that there is a  kink in the semiclassical energy at the field where the classical 
ground state is the collinear up-up-down structure, and that this kink gives rise to a locally 
linear behavior of the energy with the field when all semiclassical ground states are compared to each other for 
all fields. The magnetization curves calculated in this way for spin 1/2, 1 and 5/2 are shown to be in good agreement 
with available experimental data.
\end{abstract}

\maketitle

\section{Introduction}
 In strongly correlated electron systems quantum fluctuations are responsible for the manifestation of a variety of exotic behaviors.
 In the field of magnetic insulators, for instance, their effect can range from the stabilization of magnetic order
 to the emergence of non magnetic spin liquid phases\cite{Introduction_FM}.
 Of recent theoretical and experimental interest are the properties of frustrated magnetic insulators in external magnetic fields.
 In these systems, quantum fluctuations, which are enhanced by frustration, may lead to the presence of anomalies of the magnetization curve.
 Of specific relevance to our study are magnetization plateaus. These consist of a constant magnetization 
 at a rational value of the saturation which persists over a finite field interval.
 While plateau states break the translational symmetry of the lattice, the nature of the plateau wavefunction greatly depends on the details of the model.
 Examples include crystals of purely quantum objects such as triplet excitations in ladder systems\cite{Totsuka,Mila}, 
 crystals of more involved objects such as bound states of triplets as in the Shastry Sutherland lattice \cite{Corboz}
 or valence bond crystals 
 as identified for the $S=1/2$ Heisenberg antiferromagnet on the kagome lattice \cite{Cabra,Capponi,Nishimoto-kagome}. 
 Such plateaux are usually referred to as 'quantum' plateaux
 because the state which is stabilized has no classical analog.
 
By contrast, there are plateaux for which the magnetization pattern has a simple classical analog   consisting of a crystal 
of down pointing spins in a background of spins aligned with the magnetic field \cite{Kawamura,Chubukov-Golosov,Zhitomirsky,Penc2004,Coletta}. 
Such plateaux are sometimes referred to as 'classical' plateaux.
Given the essentially classical nature of such plateaux, it seems logical to expect that a purely semiclassical theory can be developed, and 
indeed the first prediction of a 1/3 plateau in the triangular lattice Heisenberg antiferromagnet by Chubukov and Golosov was based on semiclassical arguments\cite{Chubukov-Golosov}.
They showed, going beyond linear spin wave theory, that the $1/3$ plateau state with a $3$-sublattice up-up-down structure acquires a spin gap 
in a finite field range, and that the critical fields at which the gap closes correspond to those at which the structure stops being collinear.
 Since the seminal work of Chubukov \textit{et al.} the existence of the $1/3$ plateau was
 confirmed numerically by exact diagonalizations of finite size clusters for spin $S=1/2$ \cite{Honecker04} and $S=1$ 
 \cite{Shirata,Richter}, as well as by the coupled cluster expansion \cite{Farnell}. Moreover several experimental realizations have been
 discovered: the compound Cs$_2$CuBr$_4$, though with an orthorhombic distortion \cite{Ono1,Ono2,Tsujii,Fortune09},
 and the much closer realization of an ideal triangular lattice antiferromagnet Ba$_3$CoSb$_2$O$_9$ \cite{Shirata2,Susuki}.
 Both these compounds are relevant for the spin $S=1/2$ case. 
 Additionally we note that the compounds Ba$_3$NiSb$_2$O$_9$ and RbFe(MoO$_4$)$_2$ are other realizations of the same model but this time the on-site magnetic moment is 
respectively a spin $S=1$ \cite{Shirata,Richter} and a spin $S=5/2$ \cite{Svistov,Svistov2,Smirnov,KenzelmannLawesHarris,WhiteKenzelmann}. In all of these systems magnetization measurements report the existence of a $1/3$ plateau.

Actually, Chubukov and Golosov did not calculate the magnetization curve outside the $1/3$ plateau using a semiclassical approach.
Such a calculation has been achieved years later in the case of the square lattice antiferromagnet by Zhitomirsky and Nikuni\cite{Zhitomirsky-Nikuni}, who showed
 that a semiclassical calculation of the magnetization curve is actually possible without going beyond the linear approximation 
 if the magnetization is extracted from the derivative of the energy with respect to the field. The goal of the present paper is to show how
 this calculation can be extended to the case of the triangular lattice. This enterprise, which at first sight looks like a simple exercise,
 turned out to be far more subtle than expected, and to raise a number of interesting questions. As we shall see, the magnetization curve 
 calculated along the lines of Zhitomirsky and Nikuni
 is unphysical around the field where the classical ground state is the up-up-down state with magnetization 1/3, and 
 curing this unphysical behavior leads to an alternative semiclassical theory of the 1/3 
 magnetization plateau entirely based on energy considerations which do not require to go beyond linear order. 
 The main conclusion is that it is indeed possible to calculate the magnetization curve of the triangular lattice Heisenberg AFM including the 1/3 plateau
 within linear spin-wave theory.
 Remarkably enough, the critical fields derived by this alternative approach turn out to have the same value as those predicted by 
 Chubukov and Golosov, whose approach required to go beyond linear spin-wave theory. 

 To achieve this we will start by reminding the classical solution of the model (Sec.~\ref{sec:Classical Solution}) and
 the linear spin wave prediction for the magnetization (Sec.~\ref{sec:Spin wave expansion}).
 Then we will discuss a phenomenological theory (Sec.~\ref{sec:Plateau predictions}) which we will then put on a more
 microscopic basis in the context of a variational arguments (Sec.~\ref{sec:Variational argument}). After comparing the results 
 with available experiments (Sec.~\ref{sec:Comparison to experiments}), we will conclude with a discussion of the 
 validity and usefulness of the present results.

\section{Classical Solution} \label{sec:Classical Solution}

\begin{figure}[htbp]
 \centering
 \includegraphics[width=\columnwidth]{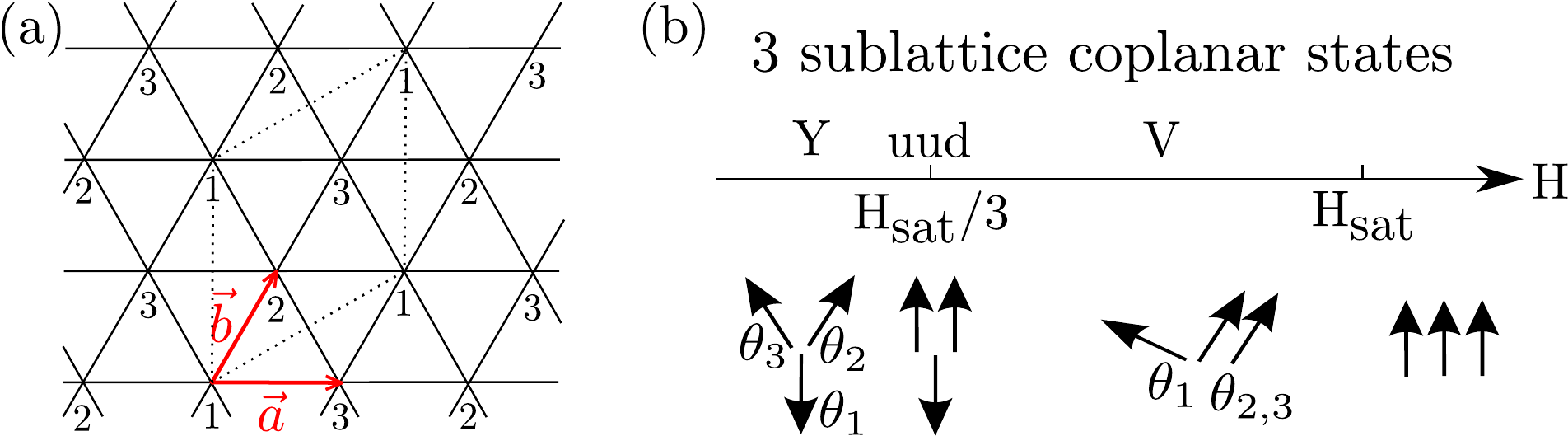}
 \caption{a) Triangular lattice and $3-$sublattice structure. The numbering indicates equivalent lattice sites.
 b) Sketch of the $3-$sublattice Y and V coplanar structures at different field values.}
 \label{fig:Coplanar structures}
\end{figure}

 The Hamiltonian of the triangular lattice Heisenberg antiferromagnet in a magnetic field is given by 
\footnote{The choice of renormalizing the bilinear spin coupling by $S^2$ and the magnetic field by $S$
formally allows to replace the quantum spin operators ${\bf S}_i/S$ by
three dimensional classical vectors of norm $1$ in the $S\rightarrow\infty$ limit. Furthermore, this choice
leads to a simple and transparent dependence in $1/S$ of the different terms of the spin wave
expansion.}
\begin{equation}\label{eq:General heisenberg model in magnetic field}
 \mathcal{H}=\frac{J}{S^2}\sum_{\langle i,j\rangle}{\bf S}_i \cdot\ {\bf S}_j-\frac{H}{S}\sum_i S_i^z\,,
\end{equation}
where the first sum is taken over all nearest neighbors of the triangular lattice [see Fig.~\ref{fig:Coplanar structures} a)].

Up to a constant the Hamiltonian (\ref{eq:General heisenberg model in magnetic field})
can be rewritten as a sum over all triangular plaquettes of the lattice as
\begin{equation}\label{eq:Hamiltonian sum over triangles}
 \mathcal{H}=\sum_{p}\frac{J}{4S^2}\left({\bf S}_{p,1}+{\bf S}_{p,2}+{\bf S}_{p,3}-\frac{{\bf H}S}{3J}\right)^2\,,
\end{equation}
with subscripts $1,2,3$ denoting the three spins belonging to the plaquette $p$.
At the classical level, when the spin operators are replaced by three dimensional vectors of norm $S$,
Eq.~(\ref{eq:Hamiltonian sum over triangles}) indicates that  the energy of the system is minimal when on all triangles of the lattice the total spin 
fulfills the constraint $({\bf S}_{p,1}+{\bf S}_{p,2}+{\bf S}_{p,3})=(S/3J){\bf H}$. 
The resulting classical ground state manifold is accidentally degenerate. For instance, both coplanar and umbrella like configurations 
minimize the classical energy. Chubukov and Golosov showed that this accidental degeneracy is lifted at $T=0$ by quantum fluctuations 
in favor of the coplanar states \cite{Chubukov-Golosov}. The $3-$sublattice coplanar states stabilized in the linear spin wave approximation can be parametrized 
by three angles measured with respect to the field direction [see Fig.~\ref{fig:Coplanar structures} b)]. They are the Y state
parametrized by $(\theta_1^Y,\theta_2^Y,\theta_3^Y)$ with $\theta_1^Y=\pi$ and $\theta_2^Y=-\theta_3^Y=\textrm{acos}{[\left(3J+H\right)/6J]}$
for $0 \leq H \leq 3J$ and the V state parametrized
by $(\theta_1^V,\theta_2^V,\theta_3^V)$ with $\theta_1^V=-\textrm{acos}{[(-27J^2+H^2)/6HJ]}$ and $\theta_2^V=\theta_3^V=\textrm{acos}{[(27J^2+H^2)/12HJ]}$
for $3J \leq H \leq 9J$.
When the field is at $1/3$ of the saturation value the Y and V states are identical to the \textit{uud} structure 
with two spins pointing along the field and one pointing down on each triangular plaquette of the lattice.

In the next section we present some basic results of the spin wave approximation for the Y and V coplanar structures.

\section{Linear-spin wave approximation}\label{sec:Spin wave expansion}

\subsection{General formalism}

The spin wave approximation consists in the bosonic reformulation of the quantum spin problem in terms
of Holstein-Primakoff (HP) particles which represent deviations from the underlying classical order
and assuming these deviations to be small compared to the size of the classical moments.
This approach is formalized in two steps: first the quantum spin Hamiltonian is rewritten by expressing the
spin operators in the local basis of the classical spin orientations denoted $(x',y',z')$.
Supposing that the coplanar Y and V structures lie in the $xz$ plane, this can be done as follows
\begin{equation}
\begin{array}{l}
  S_{{\bf R},i}^x=\cos\theta_i S_{{\bf R},i}^{x^\prime} + \sin\theta_i S_{{\bf R},i}^{z^\prime} \,, \\
  S_{{\bf R},i}^y= S_{{\bf R},i}^{y^\prime} \,, \\
  S_{{\bf R},i}^z=\cos\theta_i S_{{\bf R},i}^{z^\prime} - \sin\theta_i S_{{\bf R},i}^{x^\prime}
\end{array}
\end{equation}
where the angles $\theta_i$ parametrize the Y and V states, ${\bf R}$ is a vector of the super lattice and $i=1,2,3$ denotes the
sublattice [see Fig.~\ref{fig:Coplanar structures} a)]. In this rotated frame, the classical ground state is ferromagnetic by construction.

Secondly, deviations from the classical order are expressed in terms of the Holstein-Primakoff\cite{HolsteinPrimakoffTransformation}
representation of spin operators. To next to leading order the expressions take the form
\begin{equation}\label{eq:HP transformation TLAFM}
 \begin{array}{lll}
  S_{{\bf R},i}^{x'}&=& \displaystyle \frac{\sqrt{2S}}{2}(a_{{\bf R},i}^\dagga+a_{{\bf R},i}^\dagger)-\frac{1}{4\sqrt{2S}}\left(n_{{\bf R},i}^\dagga a_{{\bf R},i}^\dagga + a_{{\bf R},i}^\dagger n_{{\bf R},i}^\dagga \right)+\ldots\\[3mm]
  S_{{\bf R},i}^{y'}&=& \displaystyle \frac{\sqrt{2S}}{2i}(a_{{\bf R},i}^\dagga-a_{{\bf R},i}^\dagger)-\frac{1}{4i\sqrt{2S}}\left(n_{{\bf R},i}^\dagga a_{{\bf R},i}^\dagga - a_{{\bf R},i}^\dagger n_{{\bf R},i}^\dagga \right)+\ldots \\[3mm]
  S_{{\bf R},i}^{z'}&=& S-n_{{\bf R},i}^\dagga\,. \\
  \end{array}
\end{equation}
This transformation allows to rewrite the quantum Hamiltonian (\ref{eq:General heisenberg model in magnetic field})
as a sum
\begin{equation}\label{eq:HP expansion of Hamiltonian}
 \mathcal{H}=\sum_{n=0}^\infty \mathcal{H}^{(n)}\,,
\end{equation}
where $\mathcal{H}^{(n)}\propto S^{-n/2}$ contains only products of $n$
bosonic operators.
The first term of this series, $\mathcal{H}^{(0)}$, is the classical energy of the state around which fluctuations are considered.
By construction, $\mathcal{H}^{(1)}$ vanishes identically since we expand around the 3-sublattice coplanar spin configurations which are minima of 
the classical energy. $\mathcal{H}^{(2)}$ describes the single particle dynamics and all higher order terms in the expansion consist of many particle interaction
processes. 
Note that the bosonic representation is an exact mapping of the original quantum model. 
The spin wave approximation consists of a truncation scheme based on an expansion in powers of $1/S$, the inverse of the magnetic moment being the 
small expansion parameter.

\subsection{Ground state energy in the harmonic approximation}

At the harmonic approximation, which consists in truncating the expansion (\ref{eq:HP expansion of Hamiltonian}) 
to $n\leq2$, the Fourier space expression of the fluctuation Hamiltonian is given by 
\begin{equation}\label{eq:H sw}
 \mathcal{H}^{(0)}+\mathcal{H}^{(1)}+\mathcal{H}^{(2)}=  N E_\textrm{cl} + \frac{1}{2S} \sum_{\bf k} \Bigl[
 {\bf a}_{\bf k}^\dagger M_{\bf k}^\dagga(H){\bf a}_{\bf k}^\dagga-\Delta_{\bf k} \Bigr]\ ,
\end{equation}
where $E_{\textrm{cl}}=-3J/2-H^2/18J$ is the classical energy per site of the 3-sublattice coplanar states and $N$ the number of lattice sites.
Since the states considered have $3$ sites per unit cell, three distinct bosonic fields need to be introduced and thus the term ${\bf a}_{\bf k}^\dagger$ 
in Eq.~(\ref{eq:H sw}) denotes the vector $(a_{{\bf k},1}^\dagger,a_{{\bf k},2}^\dagger,a_{{\bf k},3}^\dagger,a_{{-\bf k},1}^\dagga,a_{{-\bf k},2}^\dagga,a_{{-\bf k},3}^\dagga)$.
$M_{\bf k}$ is a $6\times6$ matrix whose structure is detailed in the Appendix \ref{sec:Appendix Expressions SWT}.
The $1/S$ corrections to the classical energy are obtained by diagonalizing the fluctuation Hamiltonian (\ref{eq:H sw}) via a Bogolyubov transformation.
The diagonal representation of (\ref{eq:H sw}) consists of a sum over $3$ independent modes of free bosonic quasiparticles.
\begin{figure}[htpb]
 \centering
 \includegraphics[width=0.95\columnwidth]{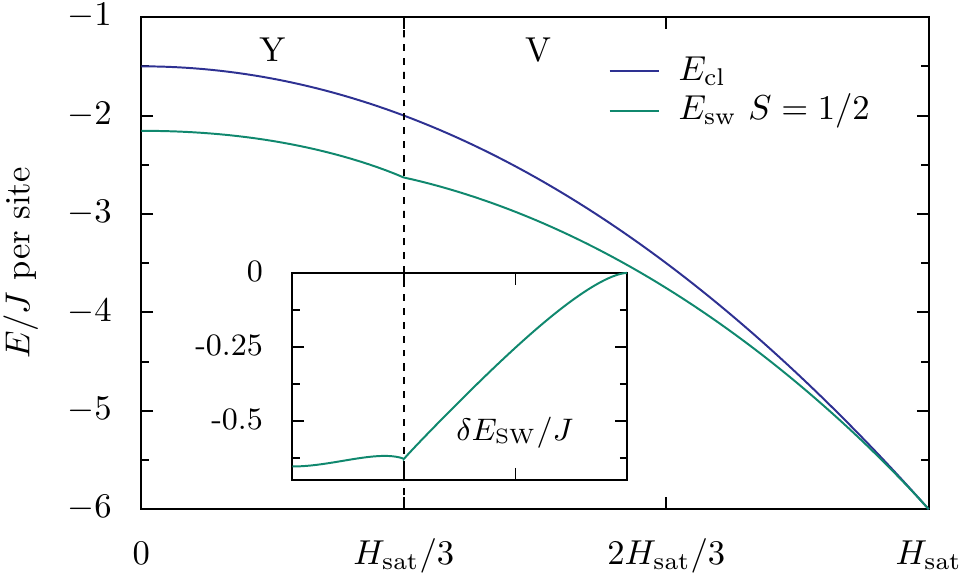}
 \caption{(Color online) Energy per site corrected by harmonic fluctuations for the coplanar Y and V type structures (green)
 and classical energy (blue). Inset: harmonic corrections to the classical energy. 
 A "kink" is visible in the energy corrected by harmonic fluctuations at the field $H_\textrm{sat}/3$.}
\label{fig:Isotropic TLAF energy 1/S 1}
\end{figure}

The ground state energy per site corrected by fluctuations at $S=1/2$ is depicted in Fig.~\ref{fig:Isotropic TLAF energy 1/S 1}.
As can be seen in the figure, the energy presents a "kink" [discontinuity in the first derivative]
at $H=H_\textrm{sat}/3$, the value of the field at which the classical ground state is the \textit{uud} state, as first noticed by Nikuni and Shiba~[\onlinecite{Shiba1}].
This cusp, present for all values of the expansion parameter $1/S$, is most pronounced for $S=1/2$.
Quantum fluctuations are responsible for the emergence of the kink in the energy, whereas the classical energy 
is differentiable (see blue curve in Fig.~\ref{fig:Isotropic TLAF energy 1/S 1}).

\subsection{Magnetization curve}

According to the Hellmann-Feynman theorem \cite{Hellmann,Feynman}, the zero temperature expression of the average magnetization per site is given by
\begin{equation}\label{eq:Magnetization HF theorem general}
 m=-\frac{1}{N}\frac{\partial E_0}{\partial H}\,,
\end{equation}
where $N$ denotes the number of lattice sites and $E_0$ is the ground state energy.
To first order in $1/S$ the magnetization can be obtained from the derivative with respect to the field of the energy corrected 
by the zero point motion\cite{Zhitomirsky-Nikuni} $E_0^\textrm{harm}$ according to
\begin{equation}\label{eq:Magnetization HF theorem}
 m=-\frac{1}{N}\frac{\partial E_0^\textrm{harm}}{\partial H}\,.
\end{equation}

\begin{figure}[htbp]
 \centering
 \includegraphics[width=0.95\columnwidth]{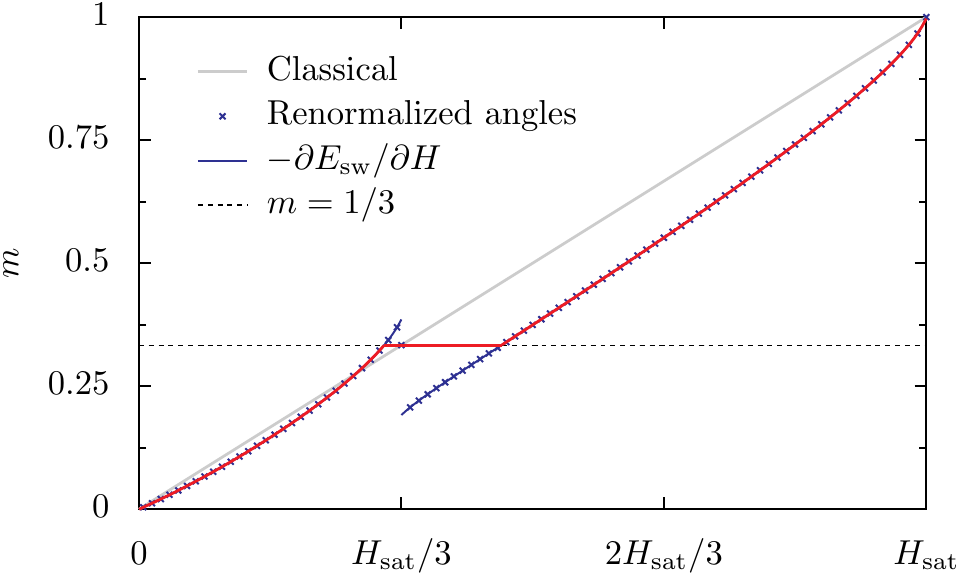}
 \caption{(Color online) Plots of the classical magnetization (black solid line) and of
 the magnetization including corrections to first order in $1/S$ for $S=1/2$ (blue curve). 
 The $1/S$ corrections to the magnetization are computed in two equivalent ways: 
 either as the derivative of the energy with respect to the magnetic field (blue curve) 
 or by direct calculation taking into account the renormalization of the spin orientations (crosses).
 The overall $1/S$ magnetization curve obtained from our phenomenological approach 
 is shown in red.}
 \label{fig:Average magnetization}
\end{figure}

The average magnetization is presented in Figure \ref{fig:Average magnetization} for $S=1/2$. 
When the $1/S$ corrections are included, the magnetization deviates from the straight line classical behavior.
As a consequence of the kink in the spin wave energy as a function of the field, the magnetization
displays a discontinuity at $H=H_\textrm{sat}/3$. Associated to the discontinuity there is a
"negative jump" in the magnetization occurring as the field is increased above $H_\textrm{sat}/3$. This
non monotonous behavior of the magnetization is of course unphysical and must be an
artifact of the harmonic truncation of the $1/S$ expansion.

\section{Phenomenological theory of magnetization}\label{sec:Plateau predictions}



Since it is known from the work of Chubukov and Golosov that there is a plateau at 1/3,
a phenomenological way to correct this unphysical aspect of the semiclassical magnetization of Fig.~\ref{fig:Average magnetization} consists 
in cutting the magnetization curve horizontally at the value $m=1/3$. 
This phenomenological approach will be put on a more systematic basis in the next section. For the moment, let us prove
that it leads to the same critical fields as Chubukov and Golosov. 

In this phenomenological approach, the critical fields are defined by the intersection between 
the magnetization curve and the line $m=1/3$. In order to extract the expressions for these critical fields one requires an analytic expression for the magnetization.
An expression for the magnetization can be extracted from Eq.~(\ref{eq:Magnetization HF theorem}). This calculation, which turns out 
to be more technical in the case of states with multiple sites per unit cell for which the explicit expression of the Bogolyubov transformation
is not known, is presented in the Appendix \ref{sec:Appendix Expressions SWT}.

Alternatively, an analytic expression of the magnetization can be obtained by computing the quantum renormalization of the 
spin orientations following the procedure of Refs.~[\onlinecite{Shiba2}] and~[\onlinecite{Zhitomirsky-Nikuni}].
In Appendix \ref{sec:Appendix Expressions SWT} it is shown that this method and the one presented in the previous paragraph
yield rigorously the same results for the magnetization.
For non collinear states the angle renormalization procedure amounts to decoupling the cubic boson term, $\mathcal{H}^{(3)}$, of the spin wave expansion which yields
an effective linear boson contribution denoted $\mathcal{H}^{(3)}_\textrm{eff}$.
The cancellation of the overall linear boson term $\mathcal{H}^{(1)}+\mathcal{H}^{(3)}_\textrm{eff}$ corresponds to a new stability condition
which is fulfilled by a new set of renormalized angles.
The renormalized spin orientations, $\tilde{\theta_i}$, are expressed for each sublattice $i$ as 
$\cos\tilde{\theta_i}=\cos{\theta_i}+c_i/S$ with the coefficients $c_i$ given by
\begin{equation}\label{eq:Angle renormalization Y}
\begin{array}{l}
\displaystyle  c_1^Y=0\,,\\
\displaystyle  c_{2,3}^Y=\cos\theta_2^Y(n_2-m_{23}-\Delta_{23})+\frac{1}{2}(-n_1+m_{21}+\Delta_{21})\,,
\end{array}
\end{equation}
for the Y state and by
\begin{equation}\label{eq:Angle renormalization V}
\begin{array}{l}
\displaystyle  c_1^V=-2\cos\theta_2^V(m_{21}+\Delta_{21})+\cos\theta_1^V n_1-\frac{3J}{H}(n_1-4n_2)\,,\\[3mm]
\displaystyle  c_{2,3}^V=-\frac{1}{2}\cos\theta_1^V(m_{21}+\Delta_{21})+\cos\theta_2^V n_2+\frac{3J}{2H}(n_1-4n_2)\,,
\end{array}
\end{equation}
for the V state, where in the above expressions we have introduced the following two body averages computed in the harmonic ground state
\begin{equation}\label{eq:Averages any H}
 n_i=\langle a_{{\bf R},i}^\dagger a_{{\bf R},i}^\dagga \rangle \,,\quad
 m_{ij}=\langle a_{{\bf R},i}^\dagger a_{{\bf  R^\prime},j}^\dagga \rangle\,, \quad
 \Delta_{ij}=\langle a_{{\bf R},i}^\dagga a_{{\bf  R^\prime},j}^\dagga \rangle\,,
\end{equation}
 with the sites $({\bf R},i)$ and $({\bf R^\prime},j)$ being nearest neighbors.

The expression of the magnetization per site in terms of the renormalized angles is
\begin{equation}\label{eq:Expression Magnetization}
 m^\textrm{Y,V}=\frac{1}{3S}\sum_{i=1}^3\cos{\tilde{\theta_i}^\textrm{Y,V}}(S-n_i)\,.
\end{equation}
Collecting all terms up to order $1/S$ in (\ref{eq:Expression Magnetization}) yields
\begin{equation}\label{eq:Magnetization}
\begin{array}{l}
\displaystyle m^Y=\frac{H}{9J}+\frac{1}{3S}\left(-2\cos\theta_2^Y(m_{23}+\Delta_{23})+m_{21}+\Delta_{21}\right)\,,\\[3mm]
\displaystyle m^V=\frac{H}{9J}\left(1-\frac{1}{S}(\Delta_{21}+m_{21})\right)\,.
\end{array}
\end{equation}
This expression of the magnetization is a function of the average quantities $n_i, m_{ij}$ and $\Delta_{ij}$ whose field dependence is presented in the Appendix 
\ref{sec:Appendix Expressions SWT}.
The magnetization (\ref{eq:Magnetization}) is reported in figure~\ref{fig:Average magnetization} and coincides with that obtained from Eq.~(\ref{eq:Magnetization HF theorem}).

Now, Chubukov and Golosov\cite{Chubukov-Golosov} showed that, to leading order in $1/S$, the fields at which the Y and V structures become collinear [i.e. that is when the renormalized 
spin orientations, measured from the field direction, tend to $(\theta_1,\theta_2,\theta_3)=(\pi,0,0)$] correspond to the critical fields at which the gaps of the renormalized 
spectra of the \textit{uud} state vanish [see Appendix \ref{sec:Appendix Expressions SWT} for more details].
Below we show that the critical fields obtained by cutting the $1/S$ magnetization curve at the value $1/3$ are the same as those predicted by Chubukov and Golosov. 
For this purpose, let us introduce the quantities $H_{c1}=3J+\alpha/S$ and $H_{c2}=3J+\beta/S$ defined such that
$m^Y(H_{c1})=m^V(H_{c2})=1/3$. Evaluating the magnetization of the Y and V states respectively at 
$H_{c1}$ and $H_{c2}$ and expanding in powers of $1/S$ gives, to lowest order,
\footnote{The superscript bar is used to emphasize that the averages $\unexpanded{\langle a_{{\bf R},i}^\dagger a_{{\bf R},i} \rangle}$,
 $\unexpanded{\langle a_{{\bf R},i}^\dagger a_{{\bf  R^\prime},j} \rangle}$ and $\unexpanded{\langle a_{{\bf R},i} a_{{\bf  R^\prime},j} \rangle}$ 
 are computed at $H=H_\textrm{sat}/3=3J$. The bar is omitted when averages are computed at different fields.}
\begin{equation}
\begin{array}{l}
 \displaystyle m^Y(H_{c1})=\frac{1}{3}+\frac{\alpha}{9JS}+\frac{1}{3S}\left(-2(\bar{m}_{23}+\bar{\Delta}_{23})+\bar{m}_{21}+\bar{\Delta}_{21}\right)\,,\\[3mm]
 \displaystyle m^V(H_{c2})=\frac{1}{3}+\frac{\beta}{9JS}-\frac{1}{3S}(\bar{\Delta}_{21}+\bar{m}_{21})\,,
\end{array}
\end{equation}
where the superscript bar denotes averages that are computed at $H=H_\textrm{sat}/3$.
Imposing $m^Y(H_{c1})=m^V(H_{c2})=1/3$ and solving for $\alpha$ and $\beta$ we obtain
 \begin{equation}\label{eq:Critical fields}
\begin{array}{l}
  \displaystyle H_{\textrm{c1}}=3J\left(1+\frac{2\bar{m}_{23}-\bar{\Delta}_{21}}{S}\right)=3J\left(1-\frac{0.084}{S}\right)\,, \\[3mm]
  \displaystyle H_{\textrm{c2}}=3J\left(1+\frac{\bar{\Delta}_{21}}{S}\right)=3J\left(1+\frac{0.215}{S}\right)\,,
\end{array}
 \end{equation}
which correspond exactly to the same $1/S$ behaviors of the critical fields predicted by Chubukov and Golosov\cite{Chubukov-Golosov}
(note that $\bar{m}_{21}=\bar{\Delta}_{23}=0$ see Appendix \ref{sec:Appendix Expressions SWT}).

So we have shown that this very simple approach to determine the plateau boundaries, which consists of cutting the average magnetization to the value $1/3$, 
produces consistent results in the large $S$ limit.
In the next section we present the formal justification of why the magnetization curve should be cut precisely at the value $m=1/3$ as well as a novel perspective
on the stabilization of the $1/3$ plateau which is based on the energetic comparison of the \textit{uud}
state with the other coplanar states.

\section{Variational theory of magnetization}\label{sec:Variational argument}

To show that cutting the magnetization at 1/3 is the right way to correct the unphysical behavior of the semiclassical magnetization
of Fig.~\ref{fig:Average magnetization}, let us first show that the existence of the kink in the energy curve corrected by harmonic fluctuations implies
that the \textit{uud} state will be stabilized over a finite field range. Our argument is the following: in the quantum Hamiltonian of the 
system (\ref{eq:General heisenberg model in magnetic field}) the total spin projection in the direction of the field is 
a conserved quantity.
Hence the energies of the eigenstates of (\ref{eq:General heisenberg model in magnetic field}) depend linearly on the field.
Now, the $1/S$ expansion of the Hamiltonian around the \textit{uud} structure preserves this property even if the expansion
is truncated at harmonic order. In the language of Holstein-Primakoff bosons
this translates into the fact that $\sum_{\bf R}\left(-n_{{\bf R},1}+n_{{\bf R},2}+n_{{\bf R},3}\right)$ 
commutes with the quadratic fluctuation Hamiltonian 
(where $1$ denotes the sublattice site with spin down and $2$ and $3$ the sublattice sites with spin up).
Therefore it is possible to determine the energy of the \textit{uud} state, which can be computed to order $1/S$
only at $H=H_{\textrm{sat}}/3$, at other values of the field according to
\begin{equation}\label{eq:uud energy}
E_{SW}^{\textrm{uud}}(H)=E_{SW}^{\textrm{uud}}(H_\textrm{sat}/3) - \frac{1}{3}\left(H-\frac{H_\textrm{sat}}{3}\right)\,,
\end{equation}
where $E_{SW}^{\textrm{uud}}(H_\textrm{sat}/3)$ is the energy per site of the \textit{uud} state corrected by
the zero point fluctuations at $H=H_\textrm{sat}/3$ and $1/3$ is the average magnetization per site
of the \textit{uud} state. 

The fact that the magnetization is strictly equal to 1/3 in the \textit{uud} state even when quantum fluctuations are
included, as anticipated in Ref.~[\onlinecite{Chubukov-Golosov}], is not completely trivial since  
the local magnetizations are no longer equal to $\pm 1/2$, but are renormalized by quantum fluctuations.
That this is true to order $1/S$ can be explicitly verified by calculating the local magnetizations at the harmonic
order, which indeed satisfy $\langle-n_{{\bf R},1}+n_{{\bf R},2}+n_{{\bf R},3}\rangle=0$.
The proof that this is true to all orders is actually even simpler. Indeed, the full quantum Hamiltonian (\ref{eq:General heisenberg model in magnetic field})
can be split into the sum of two parts $\mathcal{H}^z=J\sum_{\langle i,j\rangle}S_i^zS_j^z-H\sum_iS_i^z$ and 
$\mathcal{H}^{xy}=J\sum_{\langle i,j\rangle}S_i^xS_j^x+S_i^yS_j^y$. 
The \textit{uud} state is an eigenstate of $\mathcal{H}^z$ with magnetization equal to $1/3$ of the saturation value,
hence at the same time an eigenstate of $\sum_i S_i^z$ with eigenvalue $N/3$,
while the term $\mathcal{H}^{xy}$ is to be viewed as a perturbation to $\mathcal{H}^{z}$.
Since the commutator $[\mathcal{H}^{xy},\sum_i S_i^z]=0$ [i.e. the perturbation $\mathcal{H}^{xy}$ conserves the total spin projection in the $z$
direction] any term generated in perturbation theory starting from the \textit{uud} state has to be an eigenstate of $\sum_i S_i^z$
with the same eigenvalue $N/3$. So, the resulting eigenstate of the full Hamiltonian still has a magnetization exactly 
equal to $1/3$ of the saturation value.

\begin{figure}[htpb]
 \centering
\includegraphics[width=0.95\columnwidth]{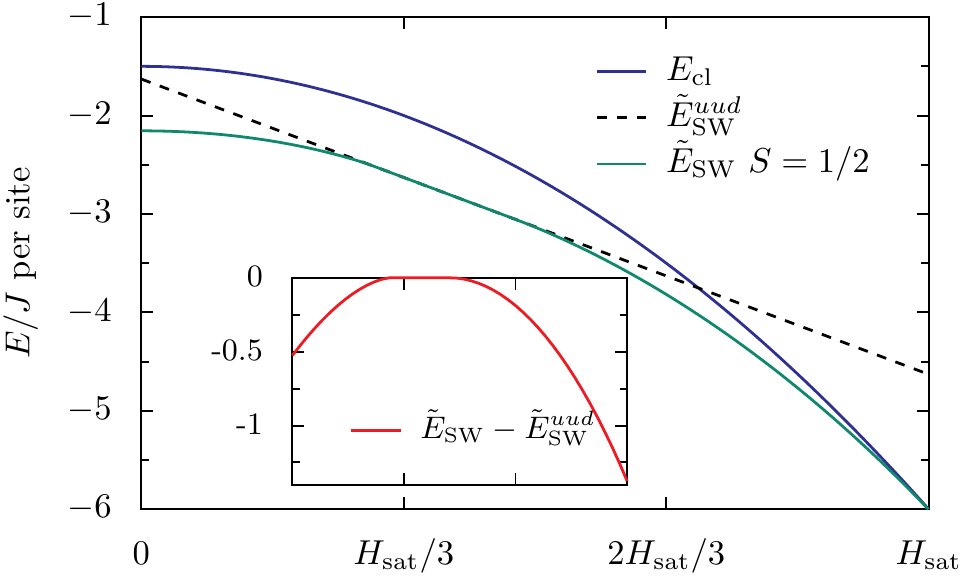}
 \caption{(Color online) The blue curve (triangles) is the classical energy and the green curve (circles) is the new 
 energy curve constructed by extrapolating linearly the energy of the different structures. 
 (Inset) Energy measured with respect to the energy of the \textit{uud} state.}
\label{fig:Isotropic TLAF energy 1/S 2}
\end{figure}

Now, since $E_{\textrm{uud}}(H_\textrm{sat}/3)$ is located at the position of the "kink" [and given the negative curvature of the energy as
a function of the field see Fig.~\ref{fig:Isotropic TLAF energy 1/S 1}] this construction indicates that in the vicinity of $H_\textrm{sat}/3$ the linear extrapolation 
of the \textit{uud} state energy (\ref{eq:uud energy}) is lower than the energy of the neighboring Y and V states. 
Thus we predict that the energy as a function of the field has a linear behavior around the kink's location $H_\textrm{sat}/3$
and that the corresponding slope is $-1/3$ [see Fig.~\ref{fig:Isotropic TLAF energy 1/S 2}]. 
This translates into a finite field interval of constant magnetization whose value is equal to $1/3$.

Simply using the linear extrapolation of the \textit{uud} state energy as a criterion for the stabilization of the plateau state
overestimates the plateau width as compared to Chubukov's result.
The reason of this overestimation is that a similar extrapolation should also be used for the neighboring non collinear Y and V states. 
Thus, we propose to compare variationally the energy of all states as follows: let $|\phi_0\rangle$ denote the ground state [i.e. the Bogolyubov vacuum]
of the harmonic fluctuation Hamiltonian around the state classically stable at $H=H_0$
then, the variational energy of this state [including harmonic fluctuations] at a different field is
given by
\begin{equation}\label{eq:Extrapolation energy of one state}
 E_0(H)=\langle \phi_0|\mathcal{H}(H_0)|\phi_0\rangle-(H-H_0) \langle \phi_0|\sum_i S_i^z/S|\phi_0\rangle.
\end{equation}
A new energy curve $\tilde{E}(H)$ is obtained by comparing, at any given field $H$, the extrapolated energies of all structures.
The resulting envelope is given by
\begin{equation}\label{eq:Etilde}
 \tilde{E}(H)=\underset{H_0}{\textrm{Min}}\left( \langle \phi_0|\mathcal{H}(H_0)|\phi_0\rangle-(H-H_0) \langle \phi_0|\sum_i S_i^z/S|\phi_0\rangle \right).
\end{equation}

In this construction we allow a given coplanar state to be stabilized at a field which is different from the one for which it is the minimum of the classical energy.
This mimics the mechanism by which quantum fluctuations renormalize the classical spin orientations.
Given that both $\langle \phi_0|\mathcal{H}(H_0)|\phi_0\rangle$ and $\langle \phi_0|\sum_i S_i^z/S|\phi_0\rangle$ are quantities which 
are the sum of a classical contribution [of order $\mathcal{O}(1)$] and of quantum corrections [of order $\mathcal{O}(1/S)$],
it can be shown that the value of $H_0$ minimizing Eq.~(\ref{eq:Etilde}) at a given $H$ is such that the difference $H-H_0$ is also of order $1/S$ 
[see Appendix \ref{sec:Appendix new energy construction} for details]. This can be understood simply by requiring that $\tilde{E}(H)$ must be equivalent to the classical energy 
in the limit $S\rightarrow\infty$, a condition that is fulfilled if the product $(H-H_0) \langle \phi_0|\sum_i S_i^z/S|\phi_0\rangle$ is a quantity which behaves as $1/S$.
Therefore, to compare the energies of states to first order $1/S$ only the classical contribution to $\langle \phi_0|\sum_i S_i^z/S|\phi_0\rangle$ needs to
be retained in Eq.~(\ref{eq:Etilde}). 

\begin{figure}[htpb]
 \centering
\includegraphics[width=0.95\columnwidth]{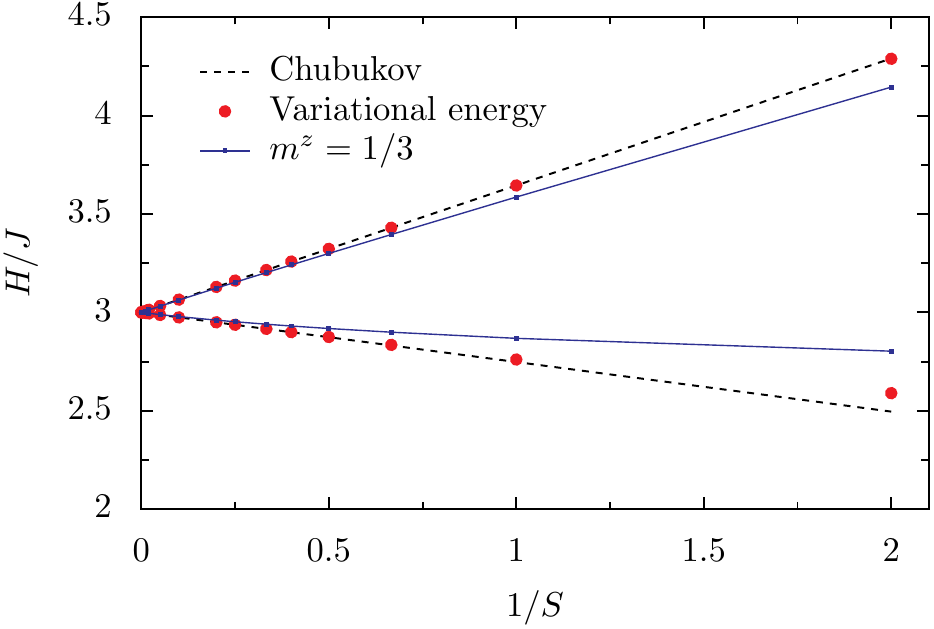}
 \caption{(Color online) Plot the $1/3$ magnetization plateau width as a function of $1/S$ estimated by different
approaches. Critical fields determined by: the condition $m=1/3$ (blue curve) and
from the variational energy construction $\tilde{E}(H)$ (red points). 
The plot also presents the extension of the lowest-order Chubukov and Golosov prediction\cite{Chubukov-Golosov} to all $S$ (dashed lines).}
\label{fig:Plateau width}
\end{figure}

In this construction the resulting energy curve $\tilde{E}(H)$ is strictly linear in the vicinity of $H_{\textrm{sat}}/3$. This behavior
corresponds to the plateau stabilization (see Fig.~\ref{fig:Isotropic TLAF energy 1/S 2}).
The plateau width obtained in this approach is reported as a function of $1/S$ in figure \ref{fig:Plateau width}.
The same plot also presents the plateau width estimates of Ref.~[\onlinecite{Chubukov-Golosov}] as well as the critical fields obtained numerically
by cutting the magnetization curve at the value $1/3$.
In all cases the agreement with Chubukov and Golosov's prediction is excellent for large $S$.

One should note that given the non trivial field dependence of the magnetization curve corrected to first order in $1/S$, solving for $H$ the equation $m(H)=1/3$
yields solutions whose expression as a series in $1/S$ includes powers of $1/S$ greater than one. This explains the discrepancy between the critical field prediction 
of this approach and that of Chubukov for large values of $1/S$ [see Fig.~\ref{fig:Plateau width}]. Nevertheless, figure~\ref{fig:Plateau width} is the numerical confirmation 
that the leading $1/S$ behaviors are the same as predicted analytically.

\section{Comparison with experiments}\label{sec:Comparison to experiments}

\begin{figure}[htbp]
\centering 
\includegraphics[width=\columnwidth]{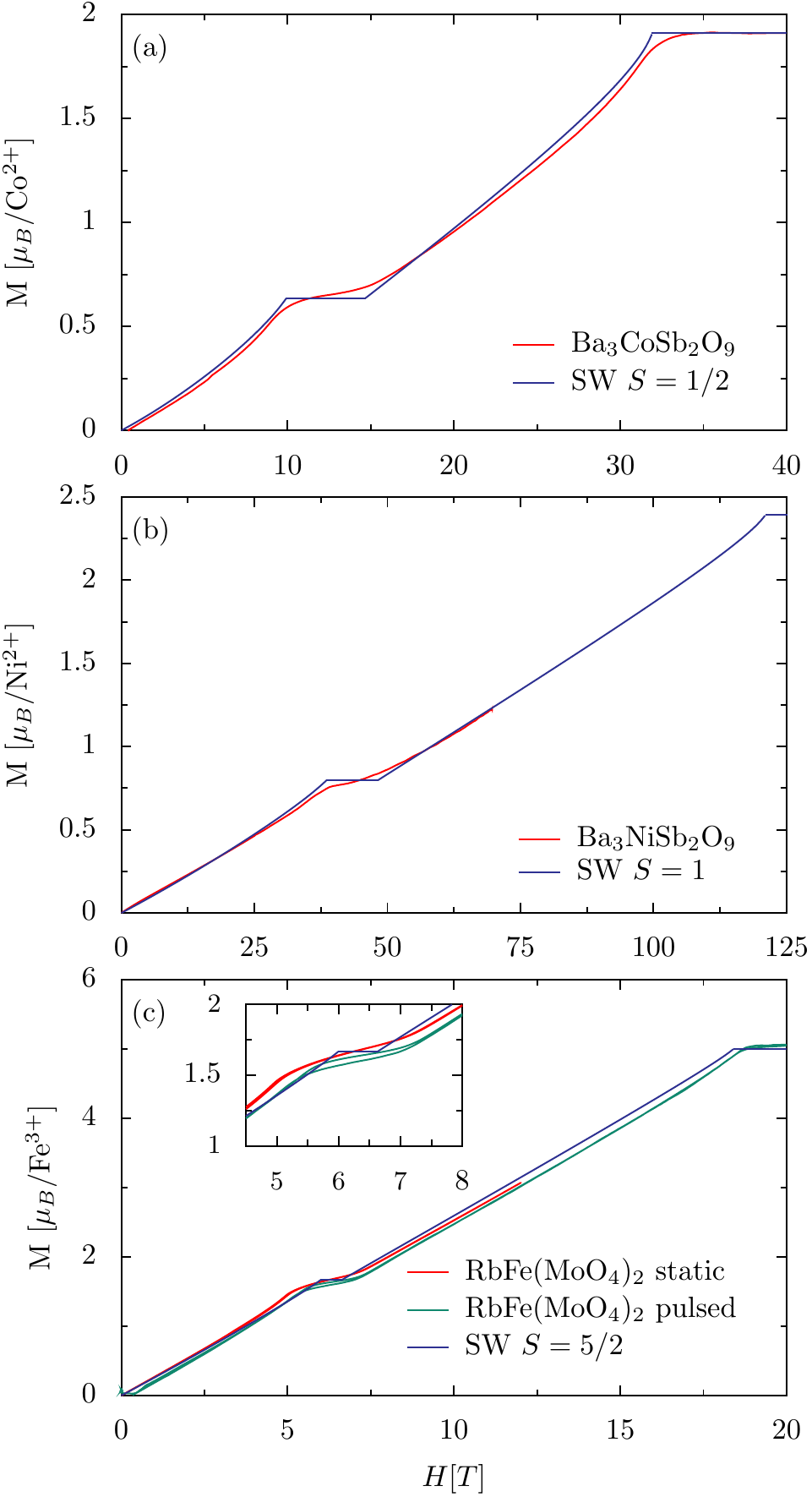}
\caption{Plot of the magnetization curve measurements for the compounds: Ba$_3$CoSb$_2$O$_9$ ($S=1/2, T=1.3 K$, powder sample) \cite{Shirata2} a), 
Ba$_3$NiSb$_2$O$_9$ ($S=1/2, T=1.3 K$, powder sample) \cite{Shirata} b), RbFe(MoO$_4$)$_2$ ($S=5/2, T=1.3 K$, pulsed field) \cite{Smirnov} 
and RbFe(MoO$_4$)$_2$ ($S=5/2, T=1.55 K$, static field) \cite{Svistov} c)
and of the $1/S$ prediction at different values of $S$.}
\label{fig:Magnetization measurements}
\end{figure}

To assess the validity of our magnetization curve construction, we compare it to recent magnetization measurements on different compounds 
which are the closest known experimental realizations of the Heisenberg model on the triangular lattice.
Figure \ref{fig:Magnetization measurements} compares the magnetization measurements for the compounds 
Ba$_3$CoSb$_2$O$_9$, Ba$_3$NiSb$_2$O$_9$ and RbFe(MoO$_4$)$_2$ [corresponding to a magnetic moment respectively of $S=1/2$, $S=1$ and $S=5/2$] to our $1/S$ prediction.

In spite of its simplicity, our theoretical prediction for the magnetization curve which consists in cutting the $1/S$ magnetization at the value $m=1/3$ yields
results in good  agreement with the experimental data both for the plateau width and position as well as for the magnetization curve away from the plateau. 
We stress however that our approach mainly provides an understanding of the plateau stabilization in the semiclassical approach.
Recent numerical studies for spin-1/2 \cite{Yamamoto14,Yamamoto15,*Sellmann15} done in the context of the magnetization process of Ba$_3$CoSb$_2$O$_9$, including an XXZ anisotropy,  are clearly more quantitative. For large spins however, our semiclassical approach is expected to be accurate.

In that respect, we note that, in spite of the larger value of the magnetic moment, the agreement of our prediction with the measurements for $S=5/2$ compound [Fig.~\ref{fig:Magnetization measurements}~c)]
is not as good as for the other compounds. 
We note however some discrepancies between the pulsed and static filed measurements in RbFe(MoO$_4$)$_2$.
Furthermore, for this compound, the saturation field is much smaller than that of the other systems. So, measured in units
of the coupling constant, the effective temperature is much larger, and temperature effects cannot be neglected. 
The general trend that the plateau is a much smaller anomaly for larger spin is nevertheless supported by the experimental data.


\section{Conclusion}\label{sec:Conclusion}

In conclusion, we have shown that a semiclassical calculation of the magnetization curve
of the Heisenberg model on the triangular lattice which includes the plateau at $1/3$ and which
is correct to order $1/S$ can be simply obtained in two steps: i) calculate the magnetization
as minus the derivative of the harmonic energy with respect to the field; ii) cut this curve
by a horizontal line at $1/3$. The justification of cutting this curve at $1/3$ relies in an 
essential way on the presence of a kink in the semiclassical energy for the field at which
the \textit{uud} state is stabilized. Thus, this simple method can be generalized to other models, 
step ii) being replaced by a cut around each point where the semiclassical energy has a kink, 
with the corresponding magnetization. 

Of course, this simple approach does not give access to all details of the magnetization
curve. In particular, it leads to cusps with finite slopes at the plateau boundaries,
whereas general arguments suggest that the transition into the plateau state should either be of the first order 
accompanied by a magnetization jump, or continuous and display a logarithmic singularity with an infinite slope since 
belonging to the same universality class of the transition into the saturated phase\cite{Takano11,Zhitomirsky-Nikuni,Gluzman,Fisher}. 
To access these details requires to go beyond the linear order in the spin wave expansion.

However, as demonstrated by the comparison with experimental data, the present theory
is quite accurate even for S=1/2, and it would presumably take experiments at very low
temperature in highly isotropic systems to actually observe significant deviations from 
the present theory, provided of course the system does not realize nonclassical
ground states on the way to polarization. Considering the difficulty in pushing spin-wave theory
beyond linear order, it is our hope that the present approach, which only relies on the 
elementary linear spin-wave theory, will be useful to both experimentalists
and theorists in the investigation of the magnetization process of frustrated quantum magnets.

\section*{ACKNOWLEDGMENTS}
We acknowledge many valuable discussions with S. Korshunov at an early stage of this project.
We are indebted to the authors of 
Refs.~[\onlinecite{Shirata}],~[\onlinecite{Shirata2}] and [\onlinecite{Smirnov}] for providing the magnetization measurements data
presented in Fig.~\ref{fig:Magnetization measurements}.
This work has been supported by the Swiss National Science Foundation and
by the Hungarian OTKA Grant No. K106047.

\appendix
\section{Spin wave theory}\label{sec:Appendix Expressions SWT}

This section presents the explicit expression of some results of the linear spin wave approximation for a generic 3-sublattice coplanar state,
as well as some aspects of the calculation to higher order referred to in the text.

\subsection{Linear spin wave approximation and $1/S$ magnetization}
The block structure of the harmonic fluctuation matrix, $M_{\bf k}$, entering Eq.~(\ref{eq:H sw})
is detailed below
\begin{equation}\label{eq:M coplanar}
M_{\bf k}=\left(
 \begin{array}{cc}
  \bar{\bar{A}}_{\bf k} & \bar{\bar{B}}_{\bf k} \\
  \bar{\bar{B}}_{\bf k} & \bar{\bar{A}}_{\bf k} 
 \end{array}\right)
\end{equation}
with
\begin{equation}\label{eq:M coplanar 2}
   \bar{\bar{A}}_{\bf k}=\left(
\begin{array}{ccc}
      A         & \gamma_{\bf k}^\star D & \gamma_{\bf k}H      \\
\gamma_{\bf k}D &       B         & \gamma_{\bf k}^\star F      \\
\gamma_{\bf k}^\star H & \gamma_{\bf k}F &        C             \\
\end{array} 
\right)
\quad \textrm{ and } \quad
   \bar{\bar{B}}_{\bf k}=\left(
\begin{array}{ccc}
       0        & \gamma_{\bf k}^\star E & \gamma_{\bf k}I      \\
\gamma_{\bf k}E &       0         & \gamma_{\bf k}^\star G      \\
\gamma_{\bf k}^\star I & \gamma_{\bf k}G &        0            
\end{array} 
\right).
\end{equation}
The coefficients entering Eq.~(\ref{eq:M coplanar 2}) are:
\begin{equation}\label{eq:Coeff M coplanar}
\begin{array}{l}
A=\left[-3J(\cos{\theta_{1,2}}+\cos{\theta_{1,3}})+H\cos\theta_1\right],\\[2mm]
B=\left[-3J(\cos{\theta_{1,2}}+\cos{\theta_{2,3}})+H\cos\theta_2\right],\\[2mm]
C=\left[-3J(\cos{\theta_{1,3}}+\cos{\theta_{2,3}})+H\cos\theta_3\right],\\[2mm]
D=J(\cos{\theta_{1,2}}+1)/2,  \quad E=J(\cos{\theta_{1,2}}-1)/2, \\[2mm]
F=J(\cos{\theta_{2,3}}+1)/2,  \quad G=J(\cos{\theta_{2,3}}-1)/2, \\[2mm]
H=J(\cos{\theta_{1,3}}+1)/2,  \quad I=J(\cos{\theta_{1,3}}-1)/2.
\end{array}
\end{equation}
where $\theta_{i,j}=\theta_i-\theta_j$ is the difference between the spin orientations on sublattices 
$i$ and $j$ (see Fig.~\ref{fig:Coplanar structures}).
The geometrical coefficient $\gamma_{\bf k}$ is given by
\begin{equation}\label{eq:Geometrical coefficients}
 \gamma_{\bf k}=\left(e^{i {\bf k}{\bf a}}+e^{-i {\bf k}{\bf b}}+e^{i {\bf k}(-\bf{a}+{\bf b}) }\right), \\[2mm]
\end{equation}
for the triangular lattice basis vectors ${\bf a}$ and ${\bf b}$ defined in Fig.~\ref{fig:Coplanar structures}a).
The additional term $\Delta_{\bf k}$ in Eq.~(\ref{eq:H sw}) is equal to the trace of $\bar{\bar{A}}_{\bf k}$, $\Delta_{\bf k}=\textrm{Tr}[\bar{\bar{A}}_{\bf k}]$.

The Bogolyubov transformation which diagonalizes (\ref{eq:H sw}) consists of a $6\times6$ momentum dependent matrix, $T_{\bf k}$,
with block structure 
\begin{equation}\label{eq:T bloc}
T_{\bf k}=\left(
 \begin{array}{cc}
  U_{\bf k} & V_{\bf k} \\
  V_{\bf k} & U_{\bf k} 
 \end{array}\right)\,.
\end{equation}
For any value of momenta, $T_{\bf k}$ simultaneously fulfills the conditions that:
i) $T^\dagger_{\bf k}M_{\bf k}T_{\bf k}$ is diagonal with doubly degenerate, real positive eigenvalues $\omega_{{\bf k},n}$
\begin{equation}\label{eq:Omega}
\begin{array}{c}
T^\dagger_{\bf k}M_{\bf k}T_{\bf k}=\Omega_{\bf k} \,,\\[3mm]
\Omega_{\bf k}=
\left(
 \begin{array}{cc}
  \omega_{\bf k} & 0 \\
  0 & \omega_{\bf k} 
 \end{array}\right)
 \,\textrm{ with }\,
\omega_{\bf k}=\left(
\begin{array}{ccc}
 \omega_{{\bf k},1} &0&0 \\
 0&\omega_{{\bf k},2} &0 \\
 0&0 &\omega_{{\bf k},3}
\end{array}
\right)\,,
 \end{array}
\end{equation}
and ii) that
\begin{equation}\label{eq:Y}
YT_{\bf k}YT_{\bf k}^\dagger=\mathbb{I}\,, \textrm{ with }
Y=\left(
 \begin{array}{cc}
  \mathbb{I} & 0 \\
  0 & -\mathbb{I}
 \end{array}\right)\,.
\end{equation}
In terms of the blocks $U_{\bf k}$ and $V_{\bf k}$, this amounts to meeting the two following requirements 
\begin{equation}\label{eq:U and V properties}
 \begin{array}{ll}
  U_{\bf k}U_{\bf k}^\dagger-V_{\bf k}V_{\bf k}^\dagger=\mathbb{I}\,,\\[2mm]
  U_{\bf k}V_{\bf k}^\dagger-V_{\bf k}U_{\bf k}^\dagger=0\,.
 \end{array}
\end{equation}
This condition (\ref{eq:U and V properties}) ensures that the Bogolyubov quasiparticles, 
which are linear combinations of the bosonic fields $a_{{\bf k},n}$ and $a_{{\bf k},n}^\dagger$,
also obey bosonic statistics.

The zero point energy per site can be expressed in terms of $T_{\bf k}$ as
\begin{equation}\label{eq:ZP energy}
 \delta E=\frac{1}{2SN}\sum_{\bf k}{\frac{1}{2}\textrm{Tr}\left[T^\dagger_{\bf k}M_{\bf k}T_{\bf k}\right]-\textrm{Tr}\left[\bar{\bar{A}}_{\bf k}\right]}\,.
\end{equation}
According to Eq.~(\ref{eq:Magnetization HF theorem}), the $1/S$ correction to the magnetization, $\delta m$, is equal to minus the derivative 
of (\ref{eq:ZP energy}) with respect to the magnetic field $H$.
Given that $\textrm{Tr}[\bar{\bar{A}}_{\bf k}]=9J$ for both the Y and V states one obtains
\begin{equation}\label{eq:derivative ZP energy}
\delta m=-\frac{1}{2SN}\sum_{\bf k}{\frac{1}{2}\textrm{Tr}\left[
T^\dagger_{\bf k}\frac{\partial M_{\bf k}}{\partial H}T_{\bf k}
+\frac{\partial T^\dagger_{\bf k}}{\partial H}M_{\bf k}T_{\bf k}
+T^\dagger_{\bf k}M_{\bf k}\frac{\partial T_{\bf k}}{\partial H}
\right]}\,.
\end{equation}
Using Eq.~(\ref{eq:Omega}), the cyclic property of the trace, and the normalization condition (\ref{eq:U and V properties}) 
one can show that the last two terms in (\ref{eq:derivative ZP energy}) vanish
\begin{equation}
\begin{array}{c}
\displaystyle\textrm{Tr}\left[\frac{\partial T^\dagger_{\bf k}}{\partial H}M_{\bf k}T_{\bf k}
+T^\dagger_{\bf k}M_{\bf k}\frac{\partial T_{\bf k}}{\partial H}\right] \\[3mm]
=\displaystyle\textrm{Tr}\left[\Omega_{\bf k}\left(\frac{\partial T^\dagger_{\bf k}}{\partial H}YT_{\bf k}Y+YT^\dagger_{\bf k}Y\frac{\partial T_{\bf k}}{\partial H}\right)\right] \\[3mm]
=\displaystyle\textrm{Tr}\left[\Omega_{\bf k}\frac{\partial}{\partial H}
\left(
\begin{array}{cc}
  U_{\bf k}U_{\bf k}^\dagger-V_{\bf k}V_{\bf k}^\dagger &  U_{\bf k}V_{\bf k}^\dagger-V_{\bf k}U_{\bf k}^\dagger \\
  U_{\bf k}V_{\bf k}^\dagger-V_{\bf k}U_{\bf k}^\dagger &  U_{\bf k}U_{\bf k}^\dagger-V_{\bf k}V_{\bf k}^\dagger \\
\end{array}\right)\right]=0\,.
\end{array}
\end{equation}
The cancellation of the terms above, which is due to the normalization conditions of the eigenvectors of $M_{\bf k}$, is 
analogous to that occurring in the Hellmann-Feynman theorem.
Hence, the $1/S$ expression for the magnetization is given by
\begin{equation}\label{eq:derivative ZP energy2}
\delta m=-\frac{1}{2SN}\sum_{\bf k}{\frac{1}{2}\textrm{Tr}\left[
T^\dagger_{\bf k}\frac{\partial M_{\bf k}}{\partial H}T_{\bf k}\right]}\,,
\end{equation}
which, given the block structure of $T_{\bf k}$, can be conveniently rewritten as 
\begin{equation}\label{eq:derivative ZP energy3}
\delta m=-\frac{1}{SN}\sum_{\bf k}\textrm{Tr}\left[\frac{\partial\bar{\bar{A}}_{\bf k}}{\partial H}V_{\bf k}V^\dagger_{\bf k}+
\frac{\partial \bar{\bar{B}}_{\bf k}}{\partial H}U_{\bf k}V^\dagger_{\bf k}\right] \,.
\end{equation}

\begin{figure}[h]
  \centering
 \includegraphics[width=0.95\columnwidth]{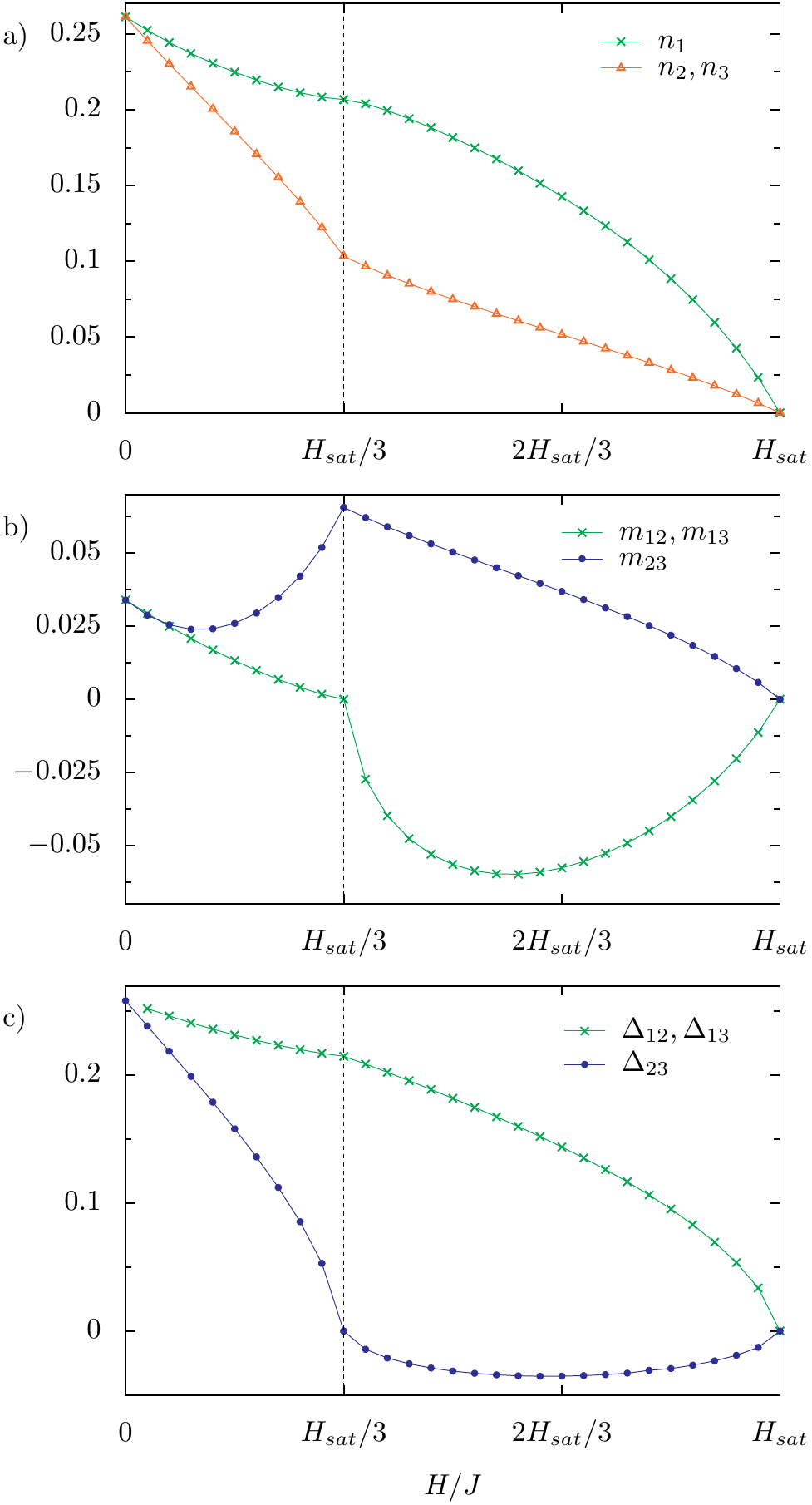} 
 \caption{(Color online) Plot, as a function of the magnetic field, of the average quantities
 $n_i=\langle a_{{\bf R},i}^\dagger a_{{\bf R},i} \rangle$ panel a), 
 $m_{ij}=\langle a_{{\bf R},i}^\dagger a_{{\bf  R^\prime},j} \rangle$ panel b),
 $\Delta_{ij}=\langle a_{{\bf R},i} a_{{\bf  R^\prime},j} \rangle$ panel c).
 We note that at $H=H_{\rm sat}/3$, i.e. for the \textit{uud} state, the quantum corrections to 
 the magnetization exactly compensate, that is $n_1-2n_2=0$.}
 \label{fig:Averages}
\end{figure}

The derivative with respect to the field of the coefficients of $\bar{\bar{A}}_{\bf k}$ and $\bar{\bar{B}}_{\bf k}$
yields
\begin{equation}\label{eq:Derivative Y}
   \frac{\partial\bar{\bar{A}}_{\bf k}^Y}{\partial H}=
   \frac{\partial\bar{\bar{B}}_{\bf k}^Y}{\partial H}=\left(
\begin{array}{ccc}
    0     & -\gamma_{\bf k}^\star/12 & -\gamma_{\bf k}/12 \\
-\gamma_{\bf k}/12 &       0         & \gamma_{\bf k}^\star\cos\theta_2^Y/3   \\
-\gamma_{\bf k}^\star/12 &  \gamma_{\bf k}\cos\theta_2^Y/3    &        0            \\
\end{array} 
\right)\,,
\end{equation}
for the Y state, and 
\begin{equation}\label{eq:Derivative V}
   \frac{\partial\bar{\bar{A}}_{\bf k}^V}{\partial H}=
   \frac{\partial\bar{\bar{B}}_{\bf k}^V}{\partial H}=\frac{H}{36J}\left(
\begin{array}{ccc}
        0             & -\gamma_{\bf k}^\star & -\gamma_{\bf k}\\
-\gamma_{\bf k}       &            0          &      0         \\
-\gamma_{\bf k}^\star &            0          &      0         \\
\end{array} 
\right)\,,
\end{equation}
for the V state.
To make contact with the alternative method to compute the $1/S$ magnetization 
presented in the main text, we note that the two body averages introduced in 
Eq.~(\ref{eq:Averages any H}) are given by the following Brillouin zone integrals
\begin{equation}\label{eq:Averages any H 2}
\begin{array}{l}
 \displaystyle n_i=\langle a_{{\bf R},i}^\dagger a_{{\bf R},i}^\dagga \rangle = \frac{3}{N}\sum_{\bf k} \left(V_{\bf k}^\dagga V^\dagger_{\bf k} \right)_{i,i} \,\,,\\
 \displaystyle m_{ij}=\langle a_{{\bf R},i}^\dagger a_{{\bf  R^\prime},j}^\dagga \rangle = \frac{1}{N}\sum_{\bf k} \gamma_{\bf k}\left(V_{\bf k}^\dagga V^\dagger_{\bf k} \right)_{i,j}\,, \\
 \displaystyle \Delta_{ij}=\langle a_{{\bf R},i}^\dagga a_{{\bf  R^\prime},j}^\dagga \rangle = \frac{1}{N}\sum_{\bf k} \gamma_{\bf k}\left(U_{\bf k}^\dagga V^\dagger_{\bf k} \right)_{i,j}\,, \\
\end{array}
\end{equation}
with the sites $({\bf R},i)$ and $({\bf R^\prime},j)$ being nearest neighbors.
The field dependence of the averages $n_i,m_{ij}$ and $\Delta_{ij}$ is reported in Fig.~\ref{fig:Averages}.
The symmetries of the Y and V structures yield $n_2=n_3$, $m_{12}=m_{13}$ , and $\Delta_{12}=\Delta_{13}$.

Injecting Eqs.~(\ref{eq:Derivative Y}), (\ref{eq:Derivative V}), and (\ref{eq:Averages any H 2}) into Eq.~(\ref{eq:derivative ZP energy3}) one recovers the
$1/S$ contribution to the magnetization presented in Eq.~(\ref{eq:Magnetization}), which is reminded below
\begin{equation}
\begin{array}{l}
\displaystyle \delta m^Y=\frac{1}{3S}\left(-2\cos\theta_2^Y(m_{23}+\Delta_{23})+m_{21}+\Delta_{21}\right)\,,\\[3mm]
\displaystyle \delta m^V=-\frac{1}{S}\frac{H}{9J}\left(\Delta_{21}+m_{21}\right)\,. 
\end{array}
\end{equation}

\subsection{Spectrum renormalization of the \textit{uud} state}
 The 3-sublattice \textit{uud} structure turns out to be classically stable at $H = H_\textrm{sat}/3$. 
 Since, according to order by disorder, collinear configurations tend to
 have a softer spectrum, hence a smaller zero-point energy \cite{Shender,*Shender2,Henley}, 
 quantum fluctuations stabilize this \textit{uud} state over a finite field
 range around $H_\textrm{sat}/3$ leading to the $1/3$ magnetization plateau \cite{Chubukov-Golosov}.
 For the specific field value $H_{\textrm{sat}}/3$, the harmonic spectrum of the $1/S$ expansion turns out to have two gapless low energy 
 modes and a higher energy gapped mode.  
 If the \textit{uud} state is to be stabilized over a given field range, it should be gapped to spin excitations.
 Chubukov and Golosov\cite{Chubukov-Golosov} showed that treating self consistently the higher order terms in the 
 spin wave expansion yields an excitation spectrum in which the two lowest bands are gapped. 
 For completeness we reproduce the main steps which lead Chubukov and Golosov to this conclusion.

 Because of collinearity, the next non vanishing term in the $1/S$ expansion around the \textit{uud} state is quartic in boson operators.
 Decoupling the quartic terms [of order $\mathcal{O}(1/S^2)$] yields an effective harmonic Hamiltonian which, up to a constant, is given by
\begin{equation}\label{eq:H sw eff}
 \mathcal{H}^{\textrm{eff}} = N E_\textrm{cl}^\textrm{uud} + \frac{1}{2S} \sum_{\bf k}
 {\bf a}_{\bf k}^\dagger \left[M_{\bf k}^\textrm{uud}(H)+\frac{1}{S}M_{\bf k}^{\textrm{eff}}\right]{\bf a}_{\bf k}^\dagga\ ,
\end{equation}
 where $M_{\bf k}^{\textrm{eff}}$ has the same block structure as $M_{\bf k}$ (\ref{eq:M coplanar}).
 The sub-blocks of $M_{\bf k}^{\textrm{eff}}$ are denoted by $\bar{\bar{A}}_{\bf k}^\textrm{eff}$ and $\bar{\bar{B}}_{\bf k}^\textrm{eff}$.
 Their expression can be obtained by replacing into Eq.~(\ref{eq:M coplanar 2}) the following coefficients
\begin{equation}\label{eq:Coeff Meffective uud}
\begin{array}{l}
A^\textrm{eff}=6J(-\bar{n}_2+\bar{\Delta}_{21})\,, \\[2mm]
B^\textrm{eff}=C^\textrm{eff}=3J(-\bar{n}_1+\bar{n}_2+\bar{\Delta}_{21}-\bar{m}_{23})\,,\\[2mm]
F^\textrm{eff}=J(\bar{m}_{23}-\bar{n}_2)\,,\quad D^\textrm{eff}=G^\textrm{eff}=H^\textrm{eff}=0\,,\\[2mm]
E^\textrm{eff}=I^\textrm{eff}=J(-\bar{\Delta}_{21}+(\bar{n}_1+\bar{n}_2)/2)\,,
\end{array}
\end{equation}
where the averages $n_i,m_{ij}$ and $\Delta_{ij}$ have been defined in Eq.~(\ref{eq:Averages any H}) [see Fig.~\ref{fig:Averages}].
The bar superscript specifies that the average quantities are computed for the field $H=H_\textrm{sat}/3$.

The contribution of the quartic terms renormalizes the harmonic spectrum opening two gaps at ${\bf k}=0$
\begin{equation}\label{eq:ChubukovGaps}
\begin{array}{l}
\displaystyle\omega_{\bf 0}^{(1)}\approx\frac{1}{S}\left[H-3J\left(1+\frac{2\bar{m}_{23}-\bar{\Delta}_{21}}{S}\right)\right]+\mathcal{O}(S^{-3})\,,\\[3mm]
\displaystyle\omega_{\bf 0}^{(2)}\approx\frac{1}{S}\left[-H+3J\left(1+\frac{\bar{\Delta}_{21}}{S}\right)\right]+\mathcal{O}(S^{-3})\,.
\end{array}
\end{equation}
 The instability of the \textit{uud} structure is resolved by determining the fields at which 
 the gaps to the first excited states close.
 To first order in $1/S$, the expression of the field values at which this takes place coincides with that given in Eq.~(\ref{eq:Critical fields}).
Ref.~[\onlinecite{Takano11}] provides a refinement of this approach which consists of a self consistent treatment 
of the decoupling of quartic terms. 

\section{Variational energy envelope}\label{sec:Appendix new energy construction}

In this Appendix we briefly mention some details of the calculation leading to the construction of a new
energy curve which supports the $1/3$ magnetization plateau in the triangular lattice Heisenberg antiferromagnet.
Let us first introduce the following notations to specify the different terms entering Eq.~(\ref{eq:Extrapolation energy of one state})
\begin{equation}\label{eq:Notations 1}
\langle\phi_0|\mathcal{H}(H_0)|\phi_0\rangle=E_{cl}(H_0)+\frac{1}{S}\delta E(H_0)\,,
\end{equation}
where $E_{cl}(H_0)$ is the classical energy at $H=H_0$ and $\delta E(H_0)/S$ the $1/S$ corrections to it.
For states different from the \textit{uud} structure, the magnetization, correct to order $1/S$, is obtained by deriving (\ref{eq:Notations 1}) with respect to the field $H_0$
\begin{equation}\label{eq:Notations 3}
\begin{array}{lll}
m(H_0)&=&\displaystyle -\frac{\partial}{\partial H_0} \left(E_{cl}(H_0)+\frac{1}{S}\delta E(H_0)\right) \\[3mm]
      &=&\displaystyle m_{cl}(H_0)+\frac{1}{S}\delta m(H_0)\,,
\end{array}
\end{equation}
where $m_{cl}(H_0)$ is the classical magnetization and $\delta m(H_0)/S$ the $1/S$ corrections to it.
Note that Eq.~(\ref{eq:Notations 3}) is meaningless at $H_\textrm{sat}/3$.
In fact, for this value of the field the harmonic energy presents a cusp and its derivative is not well defined.

The new energy curve which is proposed Eq.~(\ref{eq:Etilde}) consists of the lower envelope of all the energies defined 
in Eq.~(\ref{eq:Extrapolation energy of one state}).
As mentioned in the main text, to compare the energies of states to order $1/S$ only the classical contribution to 
$\langle \phi_0|\sum_i S_i^z/S|\phi_0\rangle$ should be retained. Thus, the minimization of (\ref{eq:Etilde}) with respect to $H_0$ 
(again for $H_0\neq H_\textrm{sat}/3$) gives
\begin{equation}\label{eq:derivative}
\begin{array}{l}
\displaystyle \frac{\partial}{\partial H_0}\left(E_{cl}(H_0)+\frac{1}{S}\delta E(H_0)\right)
+\left(1-(H-H_0)\frac{\partial}{\partial H_0}\right)m_{cl}(H_0)= 0\\
\displaystyle \Rightarrow  (H-H_0)=-\frac{1}{S}\delta m(H_0)\left(\frac{\partial m_{cl}(H_0)}{\partial H_0}\right)^{-1} \\
\displaystyle \Rightarrow  (H-H_0)=-\frac{1}{S}\frac{\delta m(H_0)}{\chi_{cl}}\,,
\end{array}
\end{equation}
where we have introduced the classical susceptibility $\chi_{cl}=\partial m(H)_{cl}/\partial H$ 
(note that $\chi_{cl}$ is a constant since the classical magnetization depends linearly with the magnetic field).
Equation (\ref{eq:derivative}) establishes that the difference $H-H_0$ which minimizes (\ref{eq:Etilde}) behaves as $1/S$.
Retaining the $1/S$ corrections of $\langle \phi_0|\sum_i S_i^z/S|\phi_0\rangle$ in the calculation would have produced a $1/S^2$ correction to 
(\ref{eq:derivative}).

Next we will show that, away from the plateau, the magnetization defined as the derivative with respect to the field of the new energy envelope
$\tilde{E}(H)$ differs from the $1/S$ magnetization (\ref{eq:Magnetization}) only by terms of order $\mathcal{O}(1/S^2)$.
For this purpose, let us compute 
\begin{equation}
  \tilde{m}(H)=-\frac{\partial \tilde{E}(H_0(H))}{\partial H}\,,
\end{equation}
where $\tilde{E}(H_0(H))$ is the new energy curve with $H_0(H)$ denoting the value of $H_0$ fulfilling (\ref{eq:Etilde}) at a given
field $H$. After derivation one obtains
\begin{equation}
 \begin{array}{lll}
  \tilde{m}(H)&=&\displaystyle-\left[\frac{\partial}{\partial H_0}\left( E_{cl}(H_0(H))+\frac{1}{S}\delta E(H_0(H))\right)\frac{\partial H_0}{\partial H}\right.\\[3mm]
              & &\displaystyle \quad\left. -\left(1-\frac{\partial H_0}{\partial H} \right)m_{cl}(H_0(H))\right.\\
              & &\displaystyle \quad \left.-(H-H_0) \frac{\partial m_{cl}(H_0(H))}{\partial H_0}\frac{\partial H_0}{\partial H}\right] \\[3mm]
              &=& m_{cl}(H_0(H))\,,
 \end{array}
\end{equation}
where we have used the first line of Eq.~(\ref{eq:derivative}) to simplify the expression.
Thus, in this construction, we are left with a new magnetization curve 
\begin{equation}\label{eq:Tilde m}
 \tilde{m}(H)=\chi_{cl}H_0(H)\,.
\end{equation}

The minimization of (\ref{eq:Etilde}) does not yield a closed form $H_0(H)$, 
however, starting from (\ref{eq:derivative}) it is straightforward to see that in 
the large $S$ limit we have
\begin{equation}\label{eq:H_0 approximate form}
 H_0=H+\frac{1}{S}\frac{\delta m(H)}{\chi_{cl}}+ \mathcal{O}(1/S^2)\,.
\end{equation}
Substituting (\ref{eq:H_0 approximate form}) into (\ref{eq:Tilde m}) produces the result announced earlier
\begin{equation}
\begin{array}{lll}
 \displaystyle\tilde{m}(H)&=&\displaystyle m_{cl}(H)+\frac{1}{S}\delta m(H)+\mathcal{O}(1/S^2)\\[3mm]
             &=&\displaystyle m(H)+\mathcal{O}(1/S^2)\,.
\end{array}
\end{equation}
So we conclude that away from the plateau, the magnetization associated with the energy curve $\tilde{E}(H)$ differs from the 
$1/S$ magnetization (\ref{eq:Magnetization}) only by terms of order $\mathcal{O}(1/S^2)$.
%
%

\bibliography{biblio} 

\end{document}